\documentstyle[12pt,epsf]{article}

\def\tb{\bar t}
\def\lsim{\mathrel{\raise.3ex\hbox{$<$\kern-.75em\lower1ex\hbox{$\sim$}}}}
\def\gsim{\mathrel{\raise.3ex\hbox{$>$\kern-.75em\lower1ex\hbox{$\sim$}}}}
\newcommand{ \slashchar }[1]{\setbox0=\hbox{$#1$}   
   \dimen0=\wd0                                     
   \setbox1=\hbox{/} \dimen1=\wd1                   
   \ifdim\dimen0>\dimen1                            
      \rlap{\hbox to \dimen0{\hfil/\hfil}}          
      #1                                            
   \else                                            
      \rlap{\hbox to \dimen1{\hfil$#1$\hfil}}       
      /                                             
   \fi}                                             %

\def\eg{{\it e.g.}}

\def\tev{\,{\rm TeV}}
\def\gev{\,{\rm GeV}}

\def\to{\rightarrow}

\def\be{\begin{equation}}
\def\ee{\end{equation}}
\def\bea{\begin{eqnarray}}
\def\eea{\end{eqnarray}}
\def\atversim#1#2{\lower0.7ex\vbox{\baselineskip\zatskip\lineskip\zatskip
  \lineskiplimit 0pt\ialign{$\matth#1\hfil##\hfil$\crcr#2\crcr\sim\crcr}}}

\renewcommand{\thefootnote}{\fnsymbol{footnote}}

\newcounter{appendixc}
\newcounter{subappendixc}[appendixc]
\newcounter{subsubappendixc}[subappendixc]

\renewcommand{\appendix}[1] {\vspace*{0.6cm}
        \refstepcounter{appendixc}
        \setcounter{figure}{0}
        \setcounter{table}{0}
        \setcounter{equation}{0}
        \renewcommand{\thefigure}{\Alph{appendixc}.\arabic{figure}}
        \renewcommand{\thetable}{\Alph{appendixc}.\arabic{table}}
        \renewcommand{\theappendixc}{\Alph{appendixc}}
        \renewcommand{\theequation}{\Alph{appendixc}.\arabic{equation}}
        \noindent{\bf Appendix \theappendixc #1}\par\vspace*{0.4cm}}

\begin{document} 

\begin{titlepage} 
\rightline{\vbox{\halign{&#\hfil\cr
&MADPH-99-1121\cr
&FERMILAB-PUB-99/221-T\cr
&hep-ph/9908236\cr
&August 1999\cr}}}
\begin{center} 
 
{\Large\bf $e^+e^- \to t\tb H$ with Non-standard Higgs Boson Couplings}

\bigskip

\normalsize 
{\bf T. Han$^{(a)}$, T. Huang$^{(b)}$, Z.-H. Lin$^{(b)}$, 
J.-X. Wang$^{(b)}$,\\ and X. Zhang$^{(b,c)}$ } \\
\vskip .3cm
$^{(a)}$Department of Physics, University of Wisconsin,\\  
Madison, WI 53706, USA\\
$^{(b)}$Institute of High Energy Physics, Academia Sinica,\\ 
Beijing, 100039, P. R. China\\
$^{(c)}$Theory Group, Fermi National Accelerator Laboratory, \\ 
P.O.Box 500, Batavia, IL 60510, USA\\
\vskip .3in

\end{center} 

\begin{abstract} 

We consider a general effective Lagrangian for couplings of a 
Higgs boson to the top-quark to dimension-six operators including 
CP violation effects. Constraints on some of them are derived
from the $Z\to b\bar b$ data. We study the process 
$e^+e^- \to t\tb H$ to probe the non-standard couplings.
We find that at a linear collider with a c.~m.~energy
$\sqrt s \sim 0.5-1.5$ TeV and a high luminosity of $10-1000$
fb$^{-1}$, these non-standard couplings may be
sensitively probed.

\end{abstract} 

\renewcommand{\thefootnote}{\arabic{footnote}} 
\end{titlepage} 


\section{Introduction}  

Given the unknown nature of electroweak symmetry breaking and the 
fact that the large top-quark mass is mysteriously close to the
electroweak scale, $m_t\approx v/\sqrt 2$ where $v\approx 246$ 
GeV is the vacuum expectation value of the Higgs field, 
it is very suggestive that the
top-quark sector may play a significant role in the electroweak
symmetry breaking \cite{hill}.
If this is the case, then the next generation of collider experiments
will have the potential to explore this fundamental physics 
associated with the Higgs and top-quark sector. 
At high energy $e^+e^-$ linear colliders, a light Higgs
boson as expected in the standard model (SM) and in Supersymmetric
(SUSY) models will be studied in detail \cite{lchiggs}. 
In connection with the top-quark
sector, the most promising process to study will be the Higgs
boson and top-quark associated production \cite{peter}
\begin{equation}
e^+ e^- \to t\bar t H.
\label{tth}
\end{equation}
By scrutinizing this process in detail, one would hope to reveal
the nature of the Higgs and top-quark interactions and hopefully
gain some insight for physics beyond the SM.

The observability of this signal over the SM backgrounds
at $e^+e^-$ colliders 
has been recently considered \cite{moretti}.
The accuracy to determine the $t\bar tH$ coupling
in the SM is also studied in detail \cite{laura}.
Possible CP-violation effects associated with the 
$t\bar tH$ vertex have been discussed \cite{pheno},
in particular, in a general two-Higgs doublet 
model \cite{jack2} and in supersymmetric models \cite{susy}.
As a model-independent approach, it may be
desirable to parameterize this class of physics by a low-energy
effective Lagrangian with unknown couplings to be determined by
experiments. Given an underlying theory, these (anomalous) 
couplings can be in principle calculated.
Such an approach has been taken in Ref.~\cite{xinmin}
in a non-linear realization of the gauge symmetry, 
and in Refs.~\cite{linear,Whisnant} in a linear realization with 
an explicit scalar (Higgs boson) field.

In this paper, we wish to take such a model-independent approach
to explore the physics in the Higgs and top-quark sector. We first
introduce a linearly realized effective Lagrangian to dimension-six
operators including CP violation. We classify them by power 
counting argument and derive constraints on some of them
from $Z\to b\bar b$ data. We calculate their effects 
at the future $e^+e^-$ linear colliders with c.~m.~energies
$\sqrt s = 0.5 - 1.5$ TeV. We then study some kinematical variables 
for process Eq.~(\ref{tth}) and construct a CP asymmetry variable.
We find that these non-standard couplings may be sensitively probed
at high energy and high luminosity $e^+e^-$ colliders.

\section{Effective Interactions of a Higgs Boson\hfill\\ 
with Top-quark}

\subsection{Effective Interactions}
 In the case of linear realization, the new physics is parameterized by
 higher dimensional operators which contain the SM fields and are
 invariant under the SM gauge group, $SU_c(3)\times SU_L(2)\times U_Y(1)$.
 Below the new physics scale $\Lambda$, the effective Lagrangian 
 can be written as
 \begin{equation}
 \label{eff}
 {\cal L}_{eff}={\cal L}_0+\frac{1}{\Lambda^2}\sum_i C_i O_i
                          +{\cal O}(\frac{1}{\Lambda^4})
			  \end{equation}
  where ${\cal L}_0$ is the SM Lagrangian.
  $O_i$ are dimension-six operators which are
  $SU_c(3)\times SU_L(2)\times U_Y(1)$ invariant and $C_i$ are constants 
which represent the coupling strengths of $O_i$ \cite{linear}. 
Recently the effective operators involving the top quark were 
reclassified and some are analyzed in Refs.~\cite{Whisnant,Renard}. 
If we assume that the new physics is of the origin associated with 
the electroweak symmetry breaking, then it is natural to identify 
the cut-off scale $\Lambda$ to be the order of ${\cal O}(4\pi v)$
and the coefficients $C_i$ the order of unity. Alternatively,
based on unitarity argument for massive quark scattering \cite{topu}, 
the scale for new physics in the top-quark sector
should be below about 3 TeV.

Following Refs.~\cite{Whisnant,Renard}, we find that there are seven 
dimension-six CP-even operators which give new contributions
to the couplings of $H$ to the top quark, 
\begin{eqnarray}
\label{O1}
O_{t1}&=&(\Phi^{\dagger}\Phi-\frac{v^2}{2})\left [\bar q_L
         t_R\widetilde\Phi
         +\widetilde\Phi^{\dagger} \bar t_R q_L\right ],\\
O_{t2}&=&i\left [\Phi^{\dagger}D_{\mu}\Phi
         -(D_{\mu}\Phi)^{\dagger}\Phi\right ]\bar t_R \gamma^{\mu}t_R,\\
O_{Dt}&=&(\bar q_L D_{\mu} t_R) D^{\mu}\widetilde\Phi
         +(D^{\mu}\widetilde\Phi)^{\dagger}(\overline{D_{\mu}t_R}q_L),\\
O_{tW\Phi}&=&\left [(\bar q_L \sigma^{\mu\nu}\tau^I t_R) \widetilde\Phi
         +\widetilde\Phi^{\dagger}(\bar t_R \sigma^{\mu\nu}\tau^I
         q_L)\right ] W^I_{\mu\nu},\\
O_{tB\Phi}&=&\left [(\bar q_L \sigma^{\mu\nu} t_R) \widetilde\Phi
         +\widetilde\Phi^{\dagger}(\bar t_R \sigma^{\mu\nu} q_L)\right ]
          B_{\mu\nu},\\
O_{\Phi q}^{(1)}&=&i\left [\Phi^{\dagger}D_{\mu}\Phi
      -(D_{\mu}\Phi)^{\dagger}\Phi\right ]\bar q_L \gamma^{\mu}q_L,\\
O_{\Phi q}^{(3)}&=&i\left [\Phi^{\dagger}\tau^I D_{\mu}\Phi
        -(D_{\mu}\Phi)^{\dagger}\tau^I\Phi\right ]\bar q_L
        \gamma^{\mu}\tau^I q_L ,
\label{O17}
\end{eqnarray}
where $\Phi$ is the Higgs doublet with 
$\widetilde\Phi = i\sigma_2 \Phi^*$,
and $\bar q_L=(\bar t_L, \bar b_L)$.
Similarly, there are seven dimension-six CP-odd 
operators \cite{Yang} which contribute to the couplings of 
$H$ to a top quark,
\begin{eqnarray}
\label{O2}
\overline {O}_{t1}&=&i(\Phi^{\dagger}\Phi-\frac{v^2}{2})
        \left [\bar q_L t_R\widetilde \Phi
        -\widetilde \Phi^{\dagger} \bar t_R q_L  \right ],\\
\overline {O}_{t2}&=&  \left [\Phi^{\dagger}D_{\mu}\Phi
      +(D_{\mu}\Phi)^{\dagger}\Phi  \right ]\bar t_R \gamma^{\mu}t_R,\\  
\overline {O}_{Dt}&=&i  \left [(\bar q_L D_{\mu} t_R) D^{\mu}\widetilde
\Phi
 -(D^{\mu}\widetilde\Phi)^{\dagger}(\overline {D_{\mu}t_R}q_L)  \right ],\\
\overline {O}_{tW\Phi}&=&i  \left [(\bar q_L \sigma^{\mu\nu}\tau^I t_R)
\widetilde\Phi
 -\widetilde\Phi^{\dagger}(\bar t_R \sigma^{\mu\nu}\tau^I q_L)  \right ]
 W^I_{\mu\nu},\\
\overline {O}_{tB\Phi}&=&i\left [(\bar q_L \sigma^{\mu\nu} t_R)
\widetilde\Phi
         -\widetilde\Phi^{\dagger}(\bar t_R \sigma^{\mu\nu} q_L)\right ]
          B_{\mu\nu},\\
\overline {O}_{\Phi q}^{(1)}&=&\left [\Phi^{\dagger}D_{\mu}\Phi
                   +(D_{\mu}\Phi)^{\dagger}\Phi\right ]
                   \bar q_L \gamma^{\mu}q_L,\\
\overline {O}_{\Phi q}^{(3)}&=&\left [\Phi^{\dagger}\tau^I D_{\mu}\Phi
                   +(D_{\mu}\Phi)^{\dagger}
                   \tau^I\Phi\right ]\bar q_L \gamma^{\mu}\tau^I q_L .
\label{Olast}
\end{eqnarray}

Operators (\ref{O1})$-$(\ref{Olast}) contribute to both the
three-point coupling $~t\bar tH~$ as well as four-point 
couplings $t\bar t H Z$ and $t\bar t H\gamma$ beyond
the SM. 
The effective Lagrangian can be viewed as a
low energy (derivative) expansion. It is thus
informative to examine their energy dependence and this
is listed in Table~\ref{one}. 
Operators $O_{t1}$ and $\overline{O}_{t1}$
give the direct corrections to the top-quark Yukawa coupling 
and is energy-independent.
Only $O_{Dt}$ and $\overline{O}_{Dt}$ contribute
to both three-point and four-point couplings 
and the three-point couplings
are quadratically dependent on energy, 
due to the nature of double-derivative couplings.
Explicit expressions in terms of physical states
for the above operators in the unitary gauge and the 
corresponding Feynman rules for the vertices 
are presented in the Appendices.
\begin{table}[thb]
\begin{center}
\begin{tabular}{|c|c|c|c|}
\hline
 & $~t\bar tH~$ & $~t\bar t H Z~$ & $~t\bar t H\gamma~$ \\ \hline
  $O_{t1},\overline{O}_{t1}$            &$1$&              &  \\ \hline
  $O_{t2}$            &        &$1/v$  &      \\ \hline
  $\overline{O}_{t2}$ &$E/v$&           &     \\ \hline
  $O_{Dt},\overline{O}_{Dt}$            &$E^2/v^2$&$E/v^2$&$E/v^2$\\ \hline
  $O_{tW\Phi},\overline{O}_{tW\Phi}$    &   &$E/v^2$&$E/v^2$\\ \hline
  $O_{tB\Phi},\overline{O}_{tB\Phi}$    &   &$E/v^2$&$E/v^2$\\ \hline
  $O_{\Phi q}^{(1)}$  &        &$1/v$&        \\ \hline
  $\overline{O}_{\Phi q}^{(1)}$&$E/v$&&        \\ \hline
  $O_{\Phi q}^{(3)}$  &        &$1/v$&        \\ \hline
  $\overline{O}_{\Phi q}^{(3)}$&$E/v$&&        \\ \hline
\end{tabular}
\end{center}
\vspace{0.1in}
\caption[]{The energy-dependence of dimension-six operators
in Eqs.~(\ref{O1})-(\ref{Olast}) for couplings
$t\bar tH$, $t\bar tHZ$ and $t\bar tH\gamma$. An overall
normalization ${v^2}/{\Lambda^2}$ has been factored out.
}
\label{one}
\end{table}


\subsection{Bounds on the Couplings}

Before we move on to discuss collider phenomenology, we first 
examine the possible constraints on those operators from the
measurement $Z\to b\bar b$. For an on-shell $Z$, 
one can write the effective vertex $Zb\bar b$ as
\begin{equation}\label{zbb}
 \Gamma_{\mu}=-i\frac{e}{4s_Wc_W} \left [\gamma_{\mu}V
-\gamma_{\mu}\gamma_5 A+\frac{1}{2m_b}(p_b-p_{\bar b})_{\mu}
S\right ],
\end{equation}
where $s_W=\sin\theta_W$; $p_b$ and $p_{\bar b}$ are the
momenta of outgoing quark and anti-quark, respectively. 
The vector and axial-vector couplings are written as
\begin{eqnarray}
V=v_b+\delta V,\quad A=a_b+\delta A,
\end{eqnarray}
where $v_b$ and $a_b$ represent the SM couplings
and $\delta V,\ \delta A$ the new physics contributions. 
To the order of $1/\Lambda^2$, the observable $R_b$ at 
the $Z$ pole is calculated to be
\begin{equation}\label{Rb}
R_b\equiv {\Gamma(Z\to b\bar b)\over \Gamma(Z\to {\rm hadrons})}
=R_b^{SM}\left[ 1+2\frac{v_b\delta V+a_b\delta A}{v_b^2+a_b^2}  
      (1-R_b^{SM})\right ],  
\end{equation}
where we have neglected the bottom quark mass.
Inversely, we have
\begin{equation}\label{Vz1}
\delta V=\delta A=
    \frac{R_b^{exp}-R_b^{SM}}{(1-R_b^{SM})R_b^{SM}}
			      \frac{v_b^2+a_b^2}{2(v_b+a_b)}.
			      \end{equation}

The SM prediction on $R_b$ and the latest experimental value
are \cite{rb}
\begin{eqnarray}\label{data1}
R_b^{SM}&=&0.2158(2),\qquad R_b^{exp}=0.21656(74).
\end{eqnarray}
If we attribute the difference as the new physics contribution,
from Eq.(\ref{Vz1}) we obtain the limit at the $1\sigma$ 
($3\sigma$) level 
\begin{equation}\label{bound1}
-3.9\times 10^{-3}~ (-8.4\times 10^{-3})<\delta V
<-5\times 10^{-5}~ (3.8\times 10^{-3}).
\end{equation}
Operators $O^{(1)}_{\Phi q}$ and $O^{(3)}_{\Phi q}$
introduced in the last section modify the
couplings at tree level. Assuming that there is no accidental 
cancellation between them, then they are calculated to be 
\begin{eqnarray}
\delta V=\delta A=-\frac{2s_Wc_W}{e}\frac{m_Z}{v}\left [
	  \frac{v^2}{\Lambda^2}C^{(1)}_{\Phi q}+\frac{v^2}
	  {\Lambda^2}C^{(3)}_{\Phi q} \right ].
\end{eqnarray}
Noting that $2s_Wc_W m_Z/ev \simeq 1$, 
we obtain the bound for each of them at the $1\sigma$ 
($3\sigma$) level as
\begin{eqnarray}\label{bound2}
5\times 10^{-5}~ (-3.8\times 10^{-3})<
\frac{v^2}{\Lambda^2}C^{(1)}_{\Phi q}=
\frac{v^2}{\Lambda^2}C^{(3)}_{\Phi q}
<3.9\times 10^{-3}~ (8.4\times 10^{-3})
\end{eqnarray}

For $O_{t1}$, $O_{t2}$, $O_{Dt}$, $O_{tW\Phi}$ and $O_{tB\Phi}$ 
they are not constrained by $R_b$ at tree level. 
However, at one-loop level they contribute to gauge
boson self-energies, and thus rather loose bounds exist \cite{Renard} 
with significant uncertainties. Bounds on them can also be studied 
from the argument of partial wave unitarity. The upper bounds 
are obtained in Ref.~\cite{Renard},
\begin{eqnarray}
|C_{t1}| & \simeq & {16\pi\over3\sqrt2}~
\left ({\Lambda\over v} \right ),\  
\label{ft2} \qquad  \qquad
|C_{t2}| \simeq 8\pi\sqrt3, \\[0.1cm]
C_{Dt} & \simeq & 10.4  \ \ \mbox{for }
C_{Dt} >0, \qquad
C_{Dt} \simeq -6.4 \ \ \mbox{for } C_{Dt} <0 , \\[0.1cm]
|C_{tW\Phi}| & \simeq & 2.5, \qquad \qquad \qquad \qquad
|C_{tB\Phi}| \simeq 2.5\ .
\end{eqnarray}

As to the new physics scale, it is plausible to envision that
$\Lambda \approx 1-3$ TeV, but we will keep $v^2/\Lambda^2$ as
a free parameter in our studies. For the convenience of
our future presentation, it is informative to see the ranges
of the unitarity bounds for $\Lambda \approx 3-1$ TeV:
\begin{eqnarray}
&& |C_{t1}|{v^2\over \Lambda^2} \simeq 1.0 - 3.0,\  
\label{ft1} \qquad  \qquad
|C_{t2}|{v^2\over \Lambda^2} \simeq 0.29 - 2.6, \\
&& C_{Dt}{v^2\over \Lambda^2} \simeq 0.07-0.63  \ \ \mbox{or } \ \
C_{Dt}{v^2\over \Lambda^2} \simeq -(0.04-0.40), \\[0.1cm]
&& |C_{tW\Phi}|{v^2\over \Lambda^2} \simeq 
|C_{tB\Phi}|{v^2\over \Lambda^2} \simeq 0.02 - 0.15\ .
\end{eqnarray}
Obviously, collider experiments have to reach a sensitivity
on these couplings below this level to be useful.

Currently, there are no significant experimental constraints on 
the CP-odd couplings involving the top-quark sector.

\section{$t\bar t H$ Production with Non-standard\hfill  \\
Couplings at Future $e^+e^-$ Colliders}

\begin{figure}[t]
\epsfysize=1.1in
\epsffile{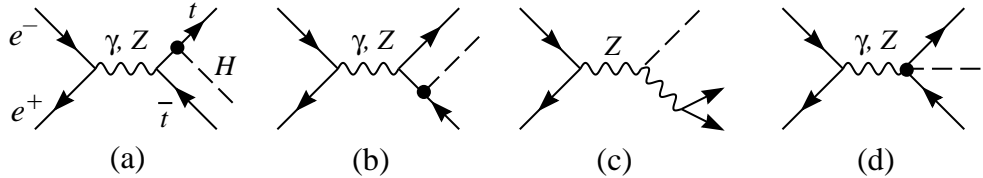}
\caption[]{Feynman diagrams for $e^+e^- \to t\bar t H$ production.
(a)-(c) are those in the SM. The dots denote the contribution
from new interactions.
\label{feyn}}
\end{figure}

One would hope to explore the new interactions presented 
in the last section at high energy colliders.
The relevant Feynman diagrams for $e^+e^- \to t\bar t H$ 
production are depicted in Fig.~\ref{feyn}, where
(a)$-$(c) are those in the SM and the dots denote the contribution
from new interactions. We evaluate all the diagrams 
including interference effects, employing a helicity
amplitude package developed in \cite{Wang}. This package
has the flexibility to include new interactions beyond
the SM. We have not included the QCD corrections to the
signal process, which are known to be positive and 
sizeable \cite{QCDtth}.

Following the power counting argument made about Table~\ref{one},
we expect that modifications to the SM prediction
from different operators would be distinctive at
high energies.
Due to the strong constraints on $O_{\Phi q}^{(1)}$ and 
$O_{\Phi q}^{(3)}$ from the $Z\to b\bar b$ measurement,
the effects of these operators at colliders will be rather small.
We will thus neglect them. 
For the purpose of illustration, we will only present
results for the operators $O_{t1}$ (energy-independent)
and $O_{Dt}$ (most sensitive to energy scale) and
hope that they are representative to the others with
similar energy-dependence based on the power-counting behavior.
For simplicity, we assume one operator to be non-zero 
at a time in our study.

\begin{figure}[thb]
\epsfysize=6in
\epsffile{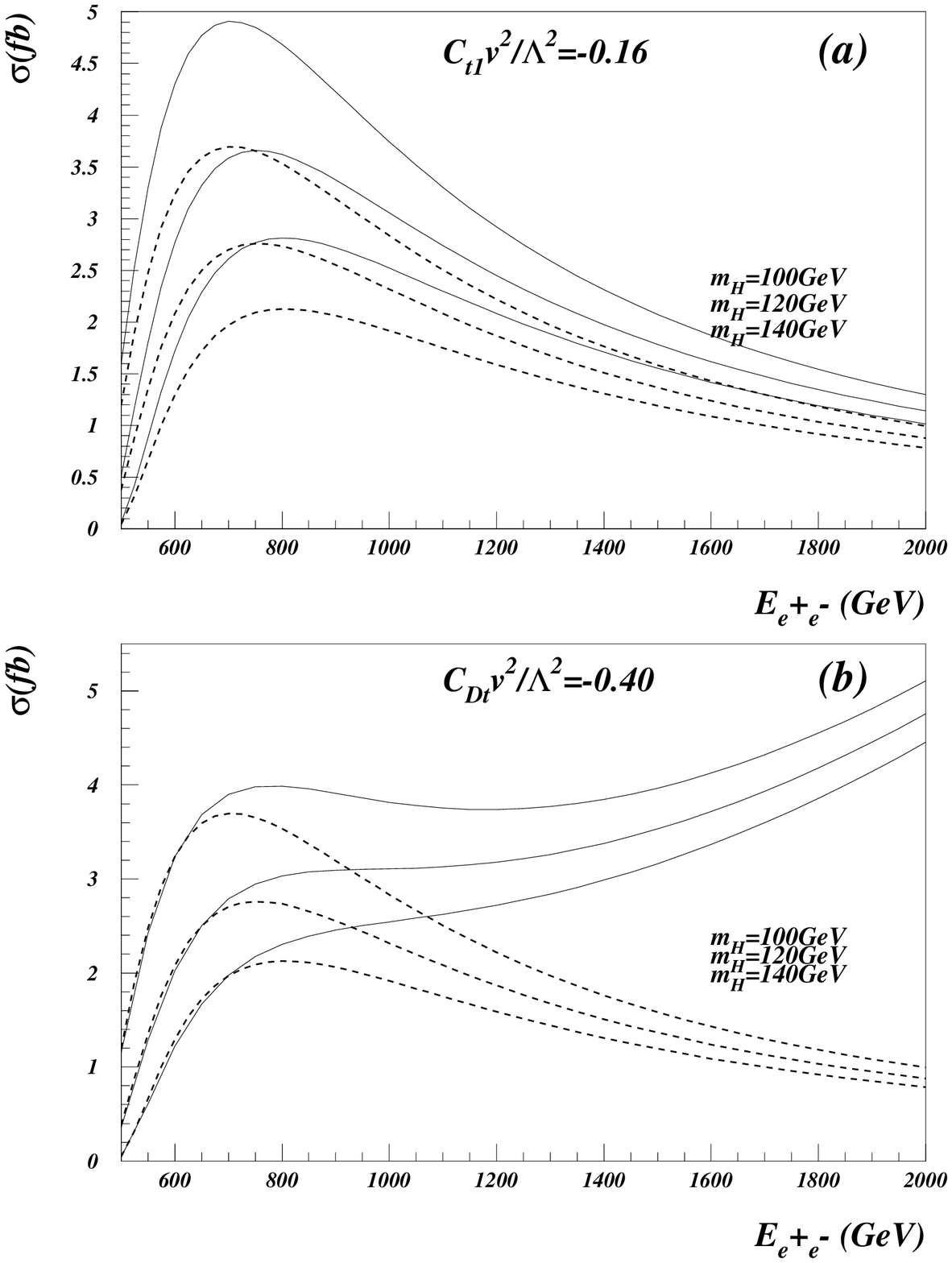}
\caption[]{Total cross section for $e^+e^- \to t\bar t H$ production
versus the $e^+e^-$ c.~m.~energy for $m_H=100$, 120 and 140 GeV, with
(a) for $O_{t1}$ and (b) for $O_{Dt}$. The dashed curves are for the
SM expectation.
\label{rts}}
\end{figure}

The production cross sections versus $\sqrt s$ are shown 
in Fig.~\ref{rts}, (a) for $C_{t1}v^2/\Lambda^2=-0.16$ 
and (b) for $C_{Dt}v^2/\Lambda^2=-0.40$
for $m_H=100,\ 120$ and 140 GeV. The dashed curves are for
the SM expectation.
As anticipated, contributions from $O_{Dt}$ become more 
significant at higher energies.

\begin{figure}[thb]
\epsfysize=6in
\epsffile{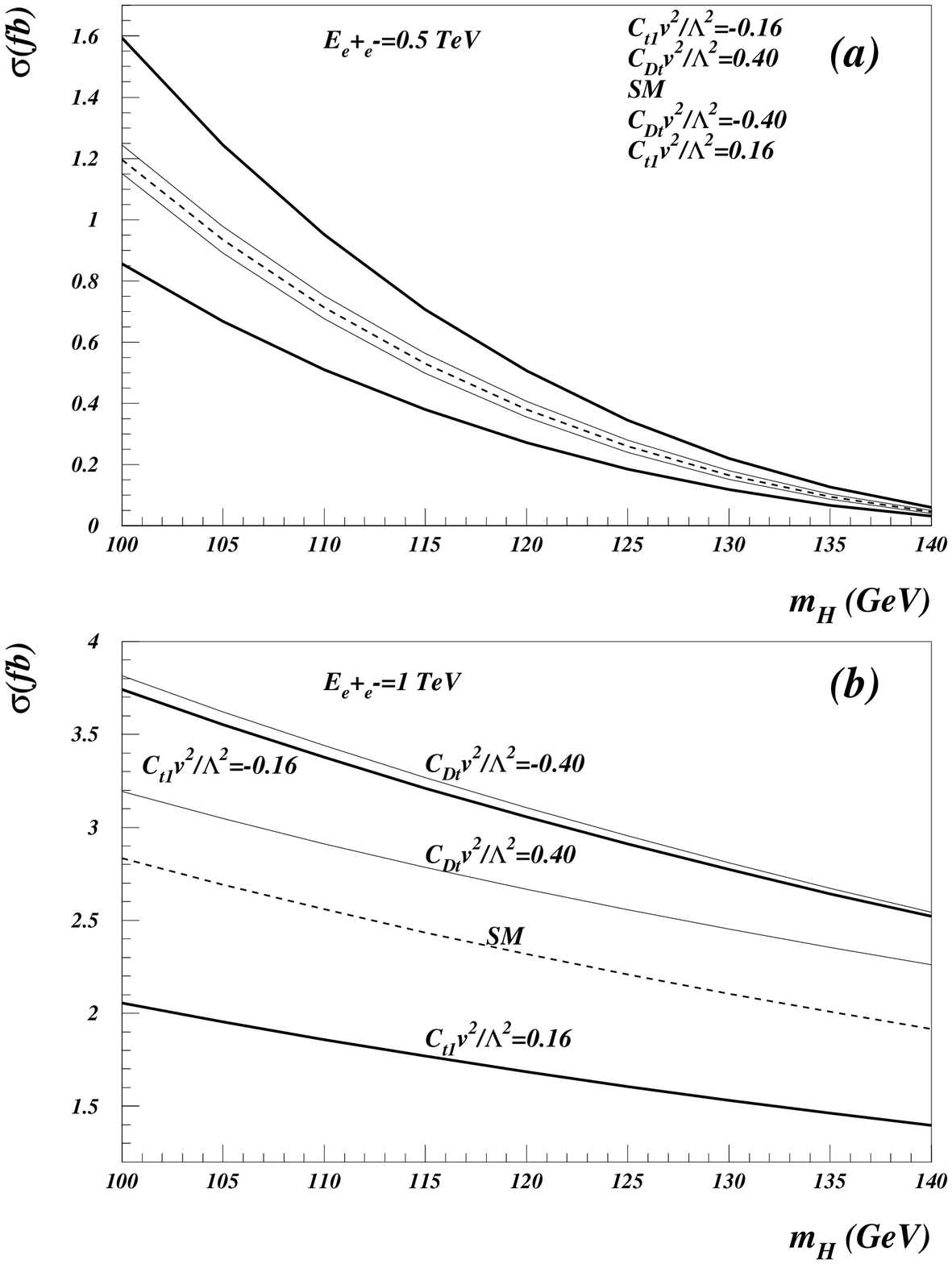}
\caption[]{Total cross section for $e^+e^- \to t\bar t H$ production
versus $m_H$ (a) for $\sqrt s=0.5$ TeV and (b) for $\sqrt s=1$ TeV.
The dashed curves are for the SM expectation.
\label{mhiggs}}
\end{figure}

We show the Higgs boson mass dependence of the cross section
in Fig.~\ref{mhiggs}, (a) for $\sqrt s=0.5$ TeV, and (b) 
for $\sqrt s=1$ TeV. A few representative values of the couplings
$C_{t1}$ and $C_{Dt}$ are illustrated. The thick solid curves are
for the operator $O_{t1}$, while the thin solid for $O_{Dt}$.

\begin{figure}[thb]
\epsfysize=7in
\epsffile{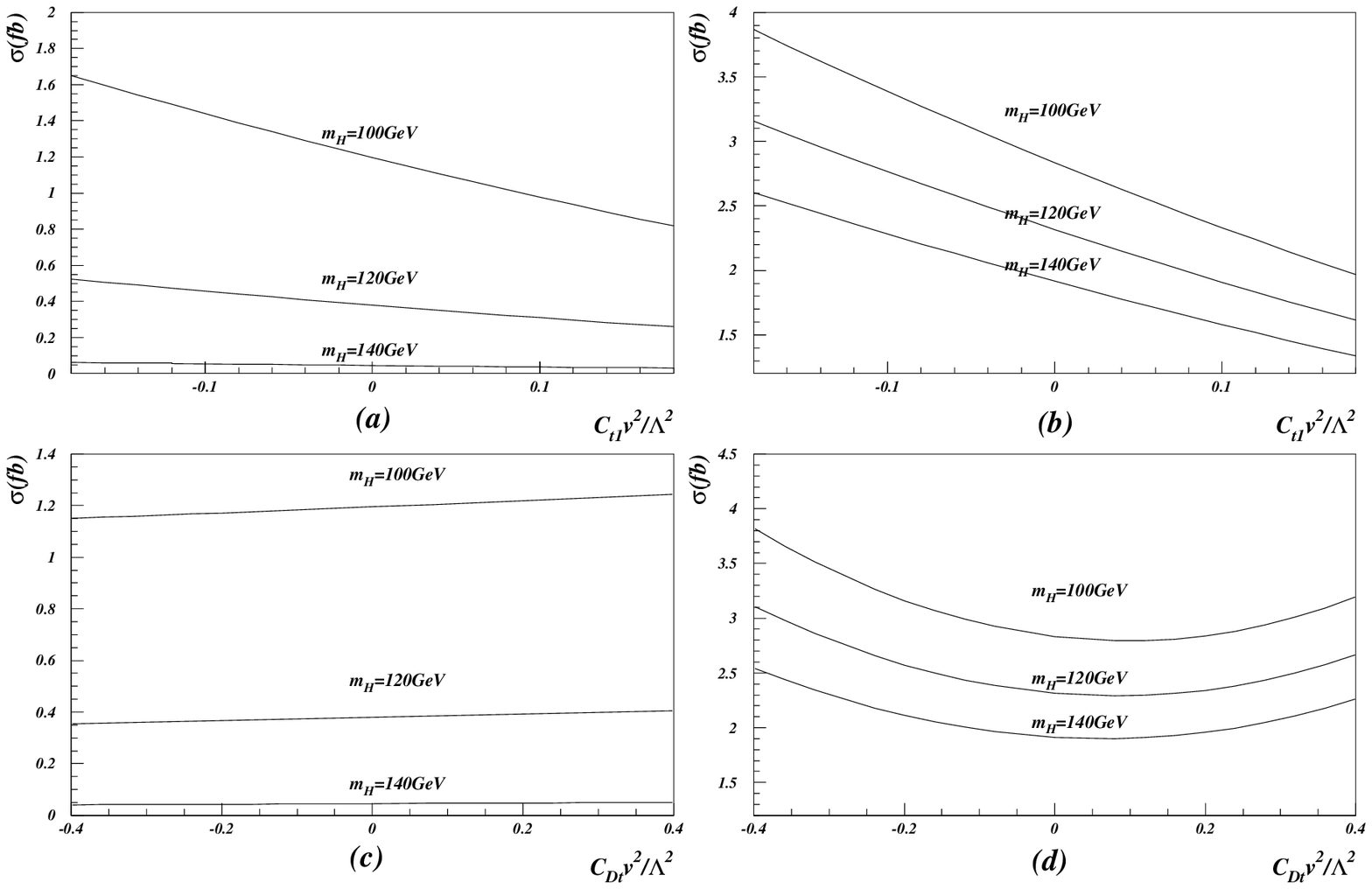}
\vskip -2cm
\caption[]{Total cross section for $e^+e^- \to t\bar t H$ production
versus the couplings 
(a) for $O_{t1}$ and $\sqrt s=0.5$ TeV ; 
(b) for $O_{t1}$ and $\sqrt s=1$ TeV  ;
(c) for $O_{Dt}$ and $\sqrt s=0.5$ TeV ;
(d) for $O_{Dt}$ and $\sqrt s=1$ TeV  
with $m_H=100,~120,~140$ GeV.
\label{coup}}
\end{figure}

It is informative to study how the cross sections change
versus the couplings, as shown in Fig.~\ref{coup}.
The four panels are 
(a) for $O_{t1}$ and $\sqrt s=0.5$ TeV; 
(b) for $O_{t1}$ and $\sqrt s=1$ TeV;
(c) for $O_{Dt}$ and $\sqrt s=0.5$ TeV;
(d) for $O_{Dt}$ and $\sqrt s=1$ TeV
with $m_H=100,~120,~140$ GeV. Due to the interference effects,
cross sections decrease as $C_{t1}$ increases and are essentially
linearly dependent upon the coupling. The effect due
to the operator $O_{Dt}$ is insignificant at $\sqrt s=0.5$ 
TeV (Fig.~\ref{coup}(c)), while at higher energies the 
contribution from $O_{Dt}$ is substantial and the quadratic
terms become quickly important (Fig.~\ref{coup}(d)).

\section{Sensitivity to the Non-standard Couplings}

To establish the sensitivity limits on the non-standard couplings
that may be probed at future linear collider experiments, 
one needs to consider the identification of the final state
from $t\bar t H$, including the
branching ratios and the detection efficiencies.
For a light Higgs boson of current interest, its leading
decay mode is $H\to b\bar b$. The branching ratio for this
mode is about $80\% \sim 50\%$ for the mass range of
$100\sim 130$ GeV.
To assure a clear signal identification, we require to identify
four $b$-jets in the final state. We assume a 65\% efficiency 
for single $b$-tagging \cite{jackson}.
As for the decays of $W^\pm$ from $t\bar t$, to effectively
increase the signal rate, we include
both the leptonic decay ($e^\pm,\mu^\pm$) \cite{moretti}
and the pure hadronic decay \cite{laura}. These amount about 85\% 
of the $t\bar t$ events. With the above event selection and 
imposing certain selective acceptance cuts, one expects to 
significantly suppress the QCD and EW background processes
$e^+e^- \to g t\bar t,\ Z t\bar t$ \cite{moretti,laura}.
We estimate an efficiency factor $\epsilon$
for detecting $e^+e^- \to t\bar t H$ to be
$$
\epsilon = 10 - 30\% ,
$$
and a factor $\epsilon '$ for reducing QCD and EW background 
to be
$$
\epsilon '=10\% 
$$
in our further evaluation.
The background cross sections for QCD ($\sigma_{QCD}$),
electroweak ($\sigma_{EW}$) and $e^+e^- \to t\bar t H$ 
in the SM ($\sigma_{SM}$) at selective energies without 
branching ratios and cuts included are listed in 
Table~\ref{two} which are consistent with that in \cite{laura}.
\begin{table}[thb]
\begin{center}
\begin{tabular}{|c|c|c|c|}
\hline
                    & $\sigma_{SM}$ &$\sigma_{EW}$  &$\sigma_{QCD}$  \\ \hline
$\sqrt s (500~\gev)$ &0.38           &0.19           &0.84            \\ \hline
$\sqrt s (1~\tev)  $ &2.32           &0.79           &1.93            \\ \hline
$\sqrt s (1.5~\tev)$ &1.36           &0.62           &1.54   \\ \hline
\end{tabular}
\end{center}
\vspace{0.1in}
\caption[]{Background cross sections
$\sigma_{SM}$, $\sigma_{EW}$ and $\sigma_{QCD}$ 
in units of fb at selective energies for $m_H=120~\gev$.
}
\label{two}
\end{table}

To estimate the luminosity ($L$) needed for probing the effects of 
the non-standard couplings, we define the significance of a signal 
rate ($S$) relative to a background rate ($B$) in terms of the
Gaussian statistics, 
\begin{eqnarray}
\label{sb}
\sigma_S = {S\over \sqrt{B}}
\end{eqnarray}
for which a signal at 95\% (99\%) confidence level (C.L.) 
corresponds to $\sigma_S=2\ (3)$. 

\subsection{ CP-even Operators}

\begin{figure}[thb]
\epsfysize=6in
\epsffile{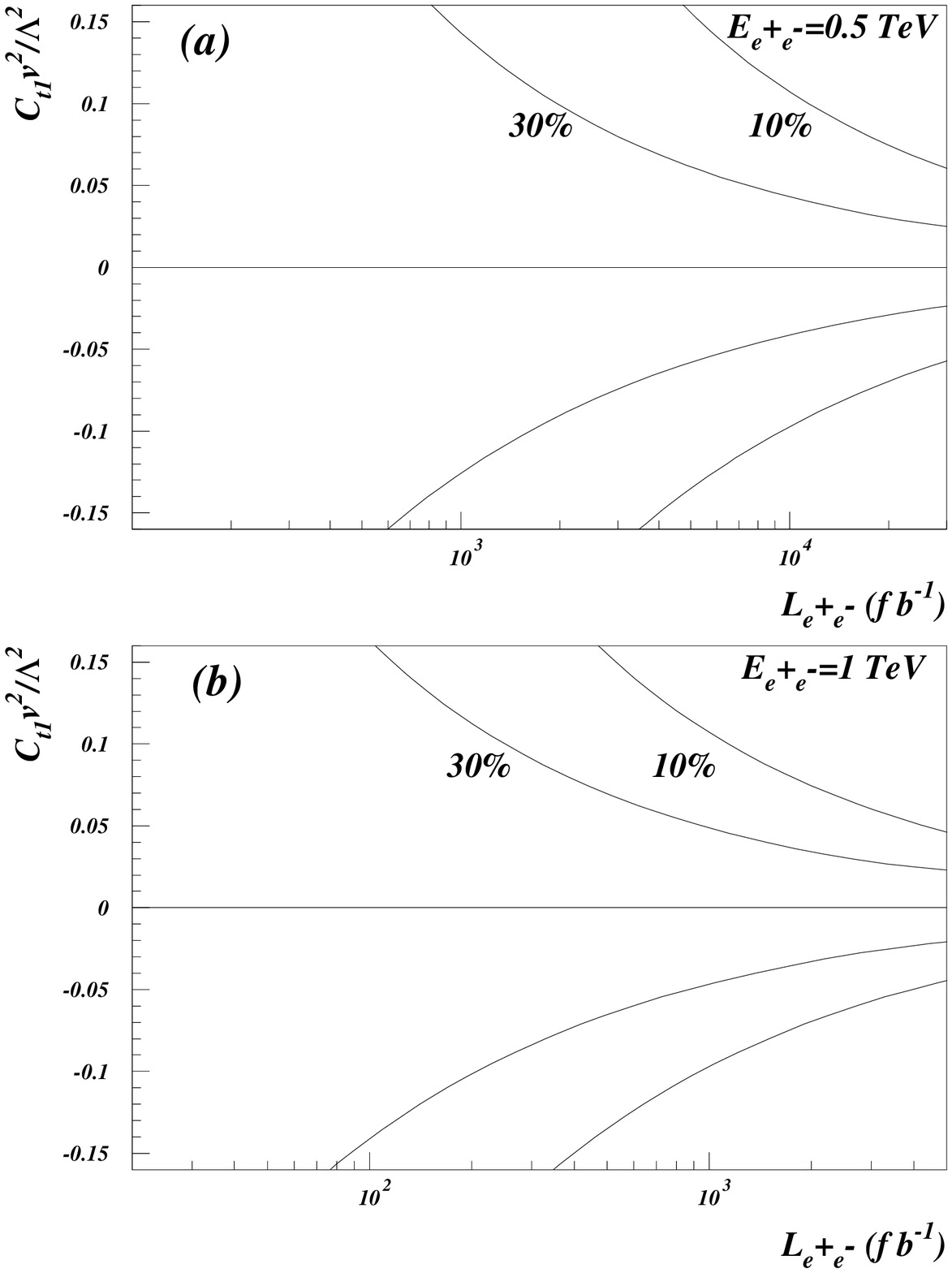}
\caption[]{Sensitivity to the anomalous couplings $O_{t1}$
versus the integrated luminosity for a 95\% confidence level limits
at (a) $\sqrt s=0.5$ TeV and (b) $\sqrt s=1$ TeV, with $m_H=120~GeV$. 
\label{lumi}}
\end{figure}

\begin{figure}[thb]
\epsfysize=6in
\epsffile{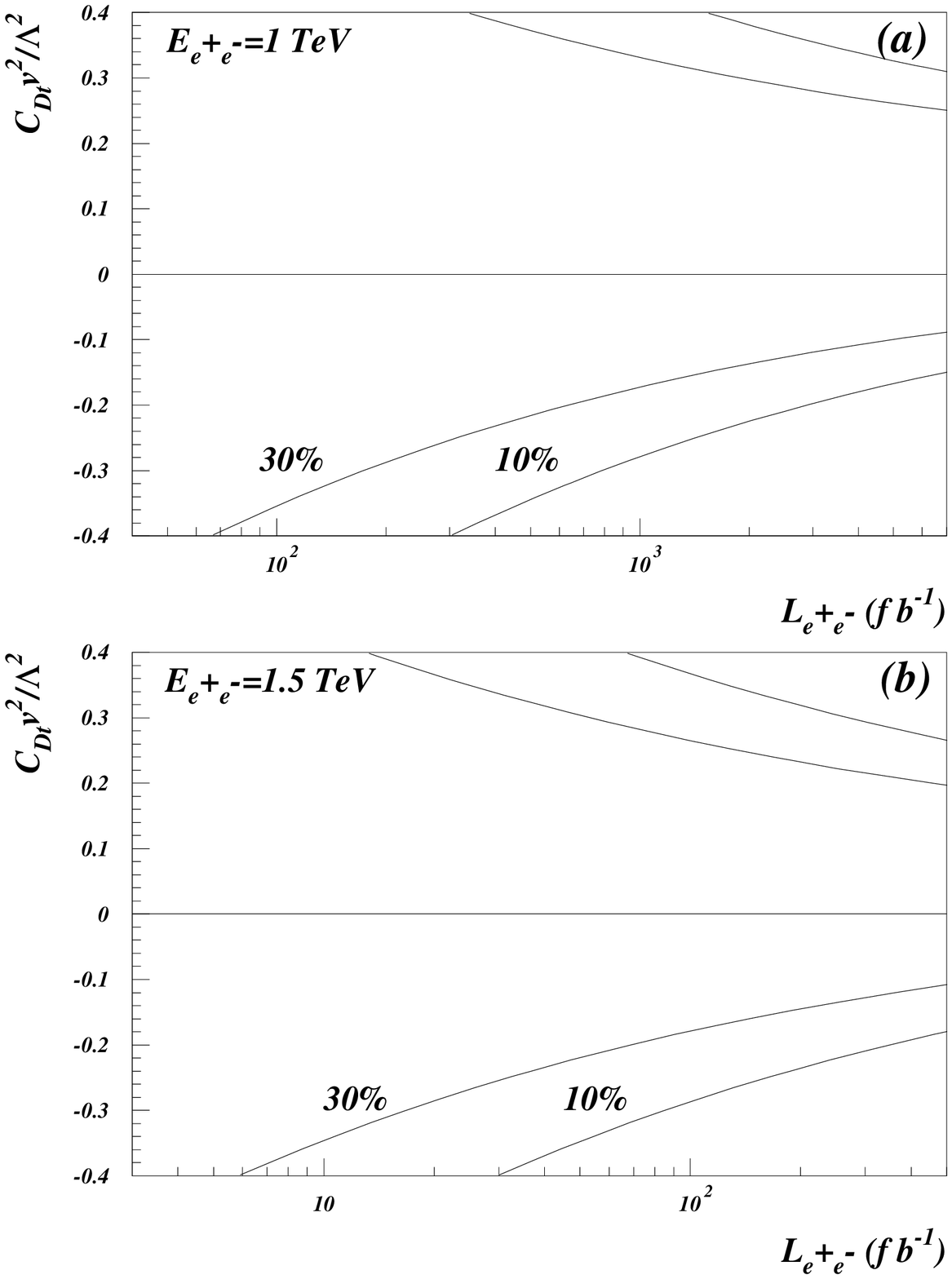}
\caption[]{Sensitivity to the anomalous couplings $O_{Dt}$
versus the integrated luminosity for a 95\% confidence level limits
at (a) $\sqrt s=1$ TeV and (b) $\sqrt s=1.5$ TeV, with $m_H=120~GeV$. 
\label{lumi2}}
\end{figure}

In the present of the CP-even operators, the $t\bar tH$
cross section ($\sigma$) would be thus modified from 
the SM expectation.
The event rates in Eq.~(\ref{sb}) are calculated as
\begin{equation}
S=L(|\sigma-\sigma_{SM}|)\epsilon\ \  {\rm and}\ \  
B=L\left [\sigma_{SM}
\epsilon+(\sigma_{QCD}+\sigma_{EW})\epsilon'\right ].
\end{equation}
We then obtain the luminosity
required for observing the effects of $O_{t1}$ at 95\% C.L. for 500
GeV and 1 TeV for $C_{t1}$ in Fig.~\ref{lumi} and for 1 TeV and 1.5 TeV
for $C_{Dt}$ in 
Fig.~\ref{lumi2}, where the two curves are for 10\% and 30\% 
of signal detection efficiency, respectively.
We see that at a 0.5 TeV collider, one would need rather 
high integrated luminosity to reach the sensitivity to
the non-standard couplings; while at a collider with a 
higher c.m. energy one can sensitively probe those
couplings with a few hundred fb$^{-1}$ luminosity.


\begin{figure}[thb]
\epsfysize=6in
\epsffile{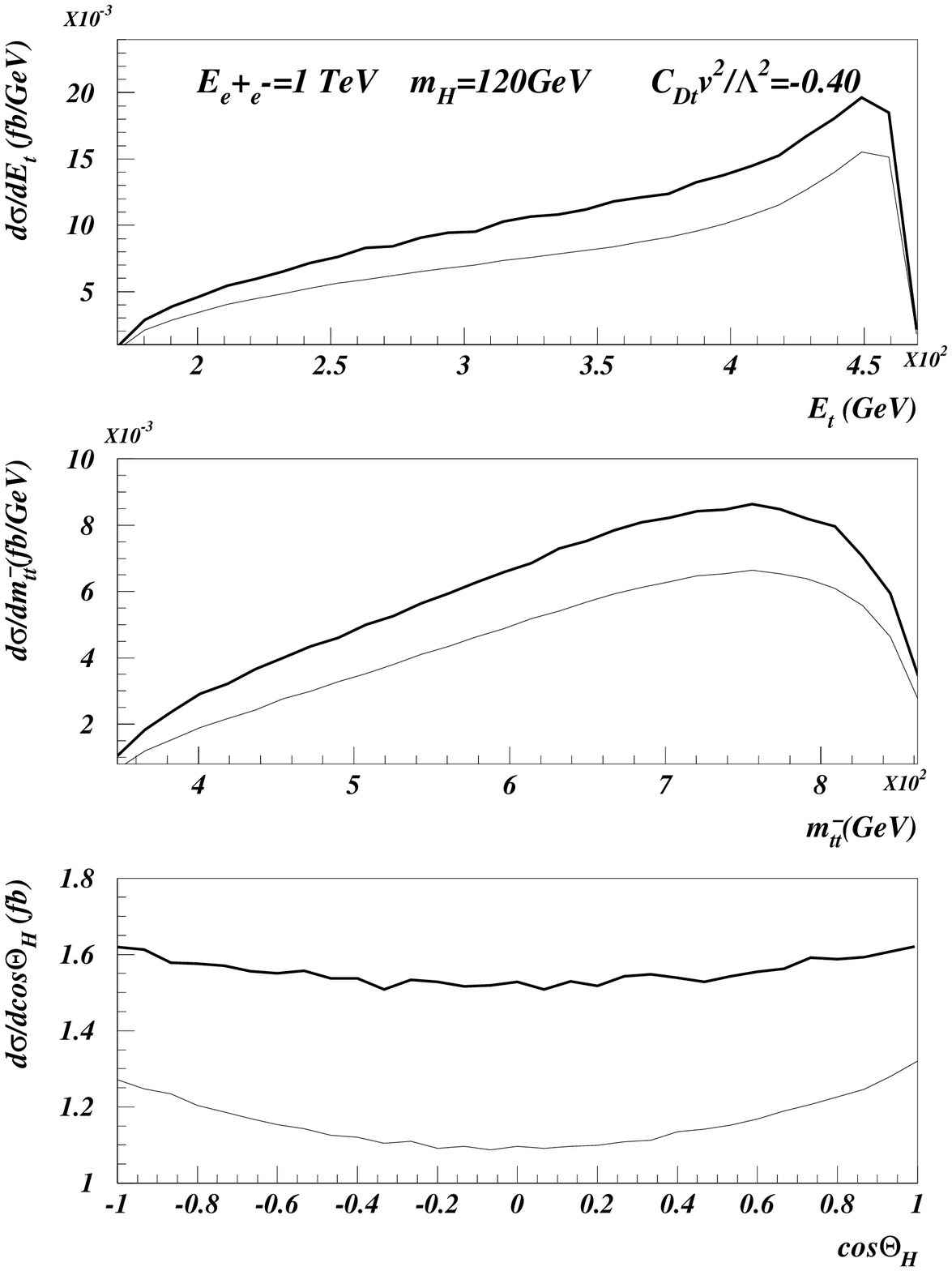}
\caption[]{Kinematical distributions for
$d\sigma/dE_t$, $d\sigma/dm_{tt}$, 
and $d\sigma/d\cos\theta_H$ with $m_H=120$ GeV, 
$C_{Dt} v^2/\Lambda^2=-0.40$
and for $\sqrt s=1$ TeV.
\label{dist}}
\end{figure}

Some kinematical distributions are discriminative for the
signal and backgrounds. For instance, 
the existence of  $O_{Dt}$ will affect the
distributions of the final state particles.
In Fig.~\ref{dist}, we plotted three distributions,
$d\sigma/dE_t$, $d\sigma/dm_{t\bar t}$
and $d\sigma/d\cos\theta_H$. Here $E_t$ and $E_H$ are the energy
of top quark and Higgs respectively, 
$m_{t\bar t}$ is the invariant mass of the $t\bar t$ system,
and $\theta_H$ is the angle of Higgs with respect to the 
electron beam direction. Non-standard couplings typically
enhance the cross section rate at higher particle 
energies and at the region of a central scattering angle.
%

\subsection{ CP-odd Operators}

\begin{figure}[thb]
\epsfysize=6in
\epsffile{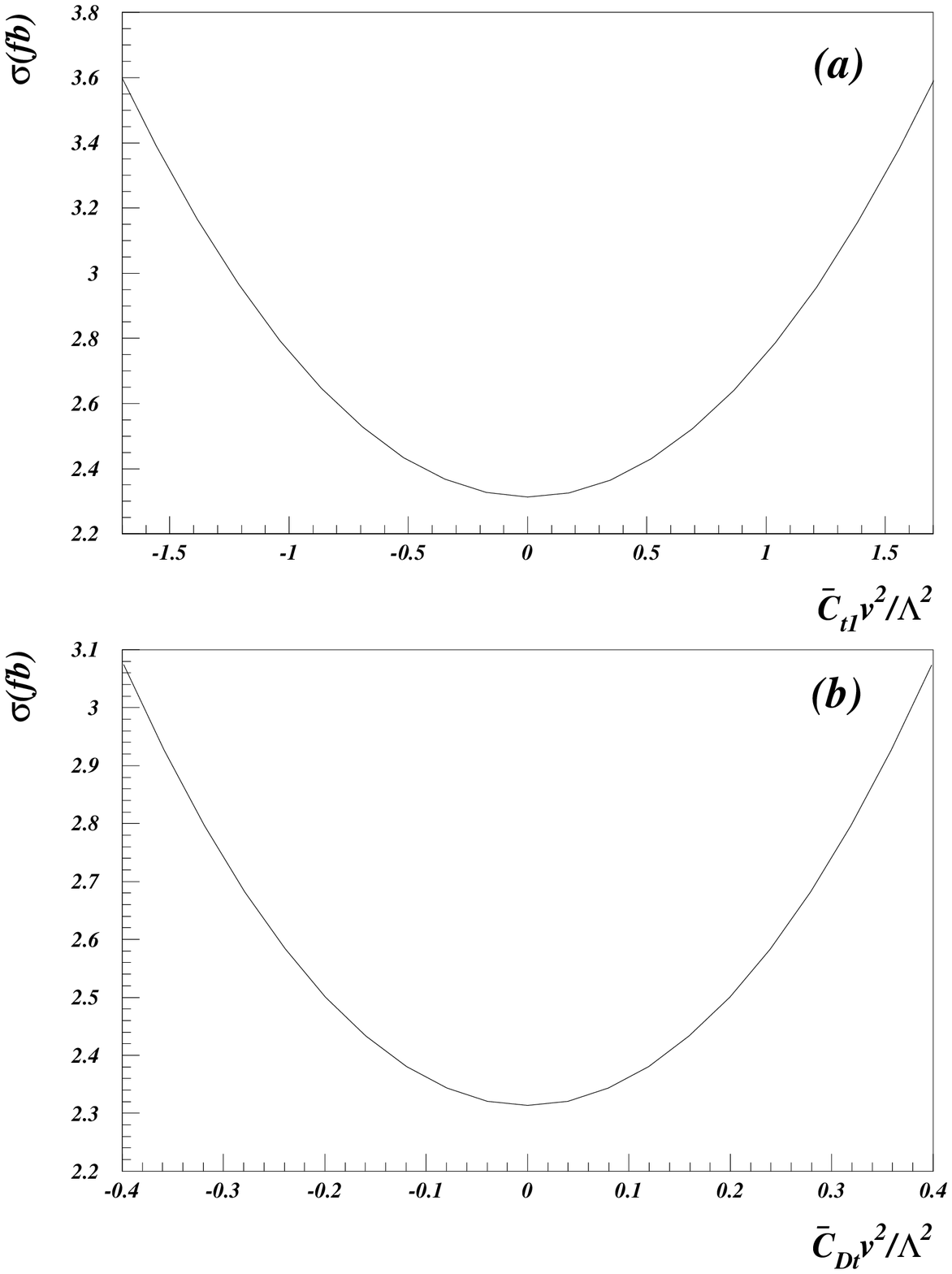}
\caption[]{
The total cross section versus CP-odd couplings for $m_H=120$ GeV, 
$\sqrt s=1$ TeV.
\label{cp1}}
\end{figure}

If there exist effective CP-odd operators beside the SM 
interaction, then CP will be violated in the Higgs and
top-quark sector. Similar to the discussion in the previous
section, one can try to observe the effects of the operators
beyond the SM expectation. 
The total cross sections versus the CP-odd operators 
$\overline {O}_{t1}$ and $\overline {O}_{Dt}$ are shown in 
Fig.~\ref{cp1} for $m_H=120$ GeV and $\sqrt s=1$ TeV.
We see that the cross sections depend on CP-odd
couplings approximately quadratically. This implies that 
the corrections to the cross sections come 
from the squared terms of matrix elements and there is
essentially no large interference between the SM and
new operators.

\begin{figure}[thb]
\epsfysize=6in
\epsffile{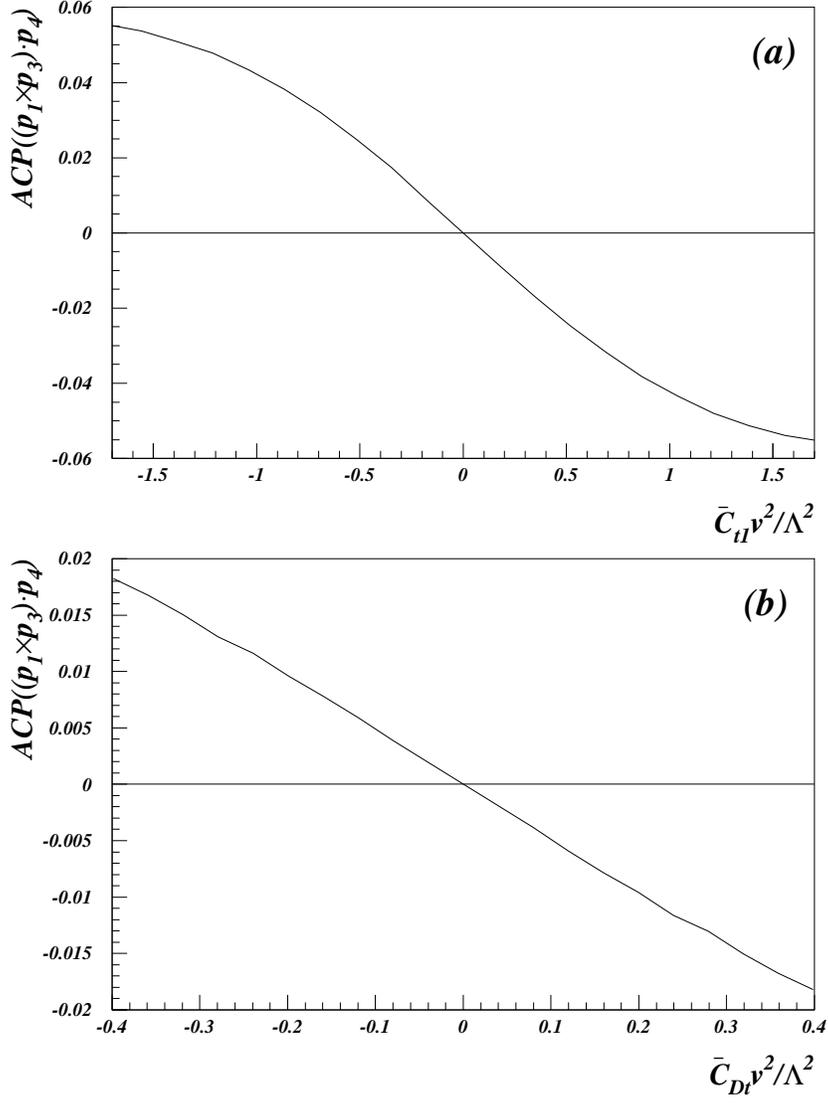}
\caption[]{
The CP asymmetry $A_{CP}$ versus CP-odd couplings for $m_H=120$ GeV, 
$\sqrt s=1$ TeV.
\label{cp2}}
\end{figure}

To unambiguously establish the observation of CP violation, one
needs to examine CP-violating observables. 
The CP-violating effect can be parameterized by a cross section
asymmetry as
\begin{eqnarray}
A_{CP}\equiv \frac{\sigma((p_1\times p_3)\bullet p_4<0)-
\sigma((p_1\times p_3)\bullet p_4>0)}
{\sigma((p_1\times p_3)\bullet p_4<0)+
\sigma((p_1\times p_3)\bullet p_4>0)}
\end{eqnarray}
where $p_1$, $p_3$ and $p_4$ are the momenta of the incoming
electron, top quark and anti-top quark, respectively.
The $1\sigma$ statistical error for $N_+$ and $N_-$ are 
$\sqrt N_+$ and $\sqrt N_-$ respectively, where
 $N_+$ is the number of the events for
 $(p_1\times p_3)\bullet p_4>0$.
  and $N_-$ is the number of the events for
  $(p_1\times p_3)\bullet p_4<0$.
  Then the error for $N_+ - N_-$ is $\sqrt { (\sqrt N_+)^2 +(\sqrt N_-)^2 }$.
  Noting that $\sqrt { (\sqrt N_+)^2 +(\sqrt N_-)^2 }=\sqrt N$,
   we get the definition of the confident level
  for two $\sigma$ as
  \begin{eqnarray}
  \label{acpsb}
  \frac{N_--N_+}{\sqrt N}=2.
  \end{eqnarray}
The asymmetry $A_{CP}$ versus the CP-odd operators 
$\overline {O}_{t1}$ and $\overline {O}_{Dt}$ is shown in 
Fig.~\ref{cp2} for $m_H=120$ GeV and $\sqrt s=1$ TeV. As one
can anticipate, $A_{CP}$ depends on the couplings linearly
since it comes from interference terms.

\begin{figure}[thb]
\epsfysize=6in
\epsffile{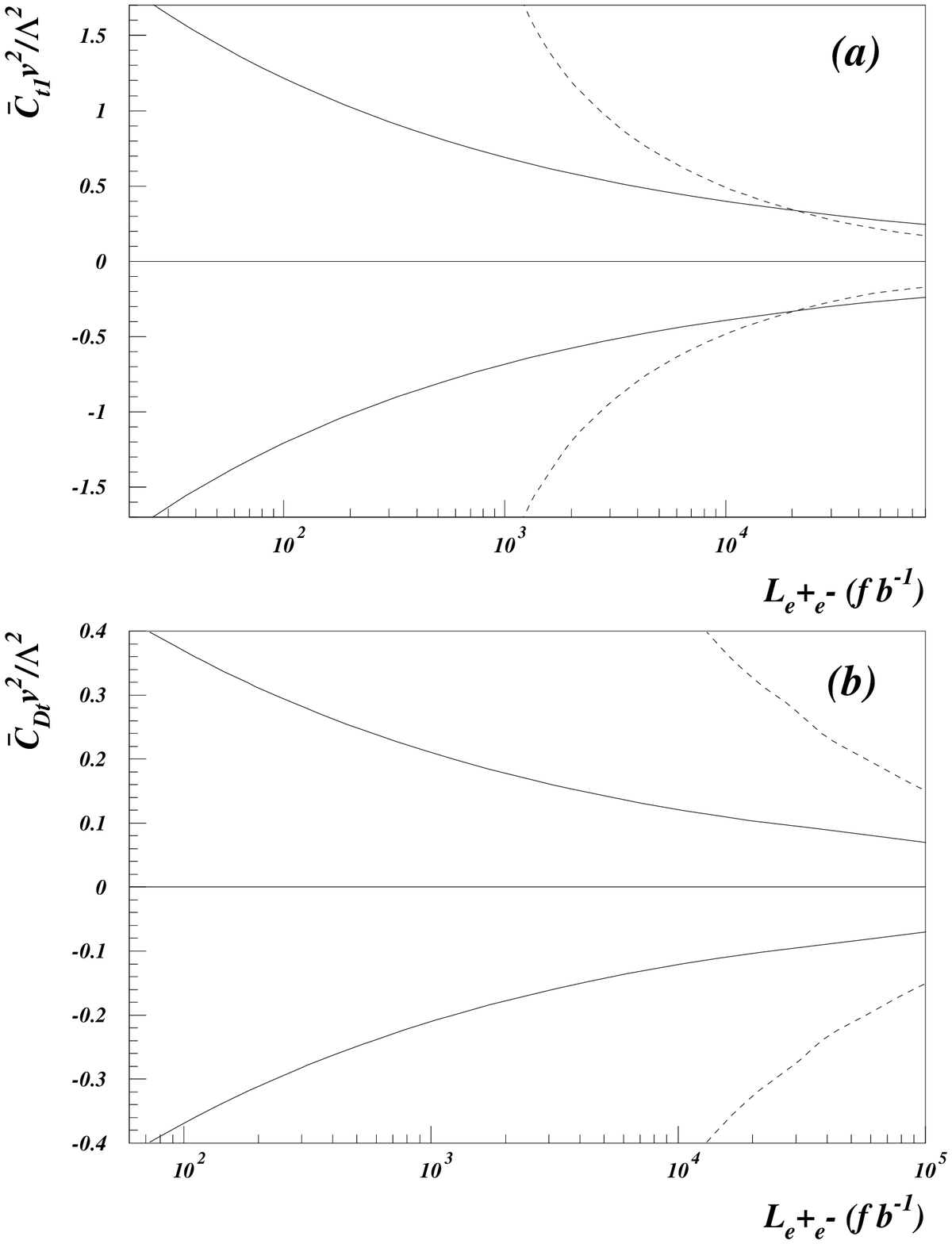}
\caption[]{
Sensitivity to the anomalous CP-odd couplings 
versus the integrated luminosity for a 95\% confidence level limits
and for $30\%$ of detection efficiency
at $\sqrt s=1$ TeV , with $m_H=120$ GeV. 
The solid line is for the total cross section and
the dash line is for the CP asymmetry $A_{CP}$.
\label{cp3}}
\end{figure}

The luminosity required for detecting the effects on the total cross sections
and $A_{CP}$ is shown in Fig.~\ref{cp3} versus CP-odd couplings with
95\% C.L for $m_H=120$ GeV and $\sqrt s=1$ TeV. The solid curves are for
the cross sections with efficiency factors $\epsilon=30\%$ and $\epsilon'=
10\%$ according to Eq.~(\ref{sb}). The dashed curves are for $A_{CP}$ with
$\epsilon=30\%$ according to Eq.~(\ref{acpsb}). 
Apparently, the effects on the total cross section due to CP-odd
operators are much stronger than that on $A_{CP}$. In other words,
the direct observation of the CP asymmetry would need much higher
luminosity to reach.


\section{Conclusions}

We have considered a general effective Lagrangian
to dimension-six operators including CP violation effects.
Constraints on some of the couplings can be derived
from the $Z\to b\bar b$ data.
We have studied the process $e^+e^- \to t\tb H$ to explore
the non-standard couplings of a Higgs boson to 
the top-quark. We found that future linear collider 
experiments should be able to probe those couplings well
below their unitarity bounds. To reach good sensitivity,
the integrated luminosities needed are about 
${\cal O}(1000,\ 100,\ 10\ {\rm fb}^{-1})$ for
$\sqrt s=0.5,\ 1.0,\ 1.5\ \tev$.

\vskip 1cm

{\it Acknowledgments}: 
We thank J.-P. Ma, Laura Reina and B.-L. Young for 
discussions. X.~Z.~would like to thank the Fermilab
theory group for hospitality during the final stage
of this work.
T. Han was supported in part by a DOE grant No. 
DE-FG02-95ER40896 and in part by the Wisconsin Alumni 
Research Foundation. 
T. Huang, Z.-H. Lin and X. Zhang were supported in part by
the NSF of China.

\appendix{~ Operators}
We work in the unitary gauge. The Higgs doublet can
be simplified as $\Phi^\dagger=(0, H+v)/\sqrt 2$.
The expressions of the seven CP-even operators in 
Eqs.~(\ref{O1})$-$(\ref{O17})
after the electroweak symmetry breaking are given by
\begin{eqnarray}
O_{t1}&=&\frac{1}{2\sqrt 2}H(H+2v)(H+v)(\bar t t),\\
O_{t2}&=&\frac{m_Z}{v}(H+v)^2 Z^{\mu}(\bar t_R \gamma_{\mu} t_R),\\
O_{Dt}&=&\frac{1}{2\sqrt 2}\partial^{\mu}H \left [\partial_{\mu}(\bar t t)
    +\bar t\gamma_5\partial_{\mu}t-(\partial_{\mu}\bar t)\gamma_5 t
    -i\frac{4}{3}g_1 B_{\mu}\bar t \gamma_5 t\right ]\nonumber\\
& & +i\frac{1}{2\sqrt 2}\frac{m_Z}{v} (H+v)Z^{\mu}\left [\bar
t\partial_{\mu}t
    -(\partial_{\mu}\bar t) t+\partial_{\mu}(\bar t\gamma_5 t)
-i\frac{4}{3}g_1 B_{\mu}\bar t t\right ]\nonumber\\
& & -i\frac{1}{2}g_2 (H+v)W_{\mu}^- \left [\bar b_L\partial^{\mu} t_R
                 -i\frac{2}{3}g_1 B^{\mu}\bar b_L  t_R\right ]\nonumber\\
& & +i\frac{1}{2}g_2 (H+v)W_{\mu}^+ \left [(\partial^{\mu} \bar t_R)b_L
                 +i\frac{2}{3}g_1 B^{\mu}\bar t_R  b_L\right ],\\   
O_{tW\Phi}&=&\frac{1}{2\sqrt 2}(H+v)(\bar t\sigma^{\mu\nu}t)
\left [W^3_{\mu\nu}-ig_2(W^+_{\mu}W^-_{\nu}-W^-_{\mu}W^+_{\nu})\right
]\nonumber\\
& & +\frac{1}{2}(H+v)(\bar b_L\sigma^{\mu\nu} t_R)  
\left [W^-_{\mu\nu}-ig_2(W^-_{\mu}W^3_{\nu}-W^3_{\mu}W^-_{\nu})\right
]\nonumber\\
& & +\frac{1}{2}(H+v)(\bar t_R\sigma^{\mu\nu} b_L)
\left [W^+_{\mu\nu}-ig_2(W^3_{\mu}W^+_{\nu}-W^+_{\mu}W^3_{\nu})\right ],\\
O_{tB\Phi}&=&\frac{1}{\sqrt 2}(H+v)(\bar t\sigma^{\mu\nu}t)B_{\mu\nu},\\
O_{\Phi q}^{(1)}&=&\frac{m_Z}{v}(H+v)^2Z_{\mu}\left [\bar
t_L\gamma^{\mu}t_L
          +\bar b_L\gamma^{\mu}b_L\right ] ,\\
O_{\Phi q}^{(3)}&=&-\frac{m_Z}{v}(H+v)^2Z_{\mu}\left [\bar
t_L\gamma^{\mu}t_L
          -\bar b_L\gamma^{\mu}b_L\right ]\nonumber\\
& &    +\frac {1}{\sqrt 2}g_2(H+v)^2\left [W^+_{\mu}\bar
t_L\gamma^{\mu}b_L   
                +W^-_{\mu}\bar b_L\gamma^{\mu}t_L\right ].
\end{eqnarray}

The expressions of the seven  CP-odd operators 
Eqs.~(\ref{O2})$-$(\ref{Olast}) after the electroweak symmetry 
breaking in the unitary gauge are given by
\begin{eqnarray}
\overline {O}_{t1}&=&\frac{1}{2\sqrt 2}H(H+2v)(H+v)(\bar ti\gamma_5 t),\\
\overline {O}_{t2}&=&(H+v)\partial^{\mu}H(\bar t_R \gamma_{\mu} t_R),\\
\overline {O}_{Dt}&=&i\frac{1}{2\sqrt 2}\partial^{\mu}H \left [\bar
t\partial_{\mu}t
                   -(\partial_{\mu}\bar t)t+\partial_{\mu}(\bar t\gamma_5
t)
                   -i\frac{4}{3}g_1 B_{\mu}\bar t t\right ]\nonumber\\
& & -\frac{1}{4\sqrt 2}g_Z (H+v)Z^{\mu}\left [\partial_{\mu}
(\bar t t)+\bar t\gamma_5\partial_{\mu}t-(\partial_{\mu}\bar t)\gamma_5 t
-i\frac{4}{3}g_1 B_{\mu}\bar t \gamma_5 t\right ]\nonumber\\
& & +\frac{1}{2}g_2 (H+v)W_{\mu}^- \left [\bar b_L\partial^{\mu} t_R
                 -i\frac{2}{3}g_1 B^{\mu}\bar b_L  t_R\right ]\nonumber\\
& & +\frac{1}{2}g_2 (H+v)W_{\mu}^+ \left [(\partial^{\mu} \bar t_R)b_L  
                 +i\frac{2}{3}g_1 B^{\mu}\bar t_R  b_L\right ],\\
\overline {O}_{tW\Phi}&=&i\frac{1}{2\sqrt 2}(H+v)(\bar
t\sigma^{\mu\nu}\gamma_5 t)  
      \left [W^3_{\mu\nu}-ig_2(W^+_{\mu}W^-_{\nu}
      -W^-_{\mu}W^+_{\nu})\right ]\nonumber\\
& & +i\frac{1}{2}(H+v)(\bar b_L\sigma^{\mu\nu} t_R)
      \left [W^-_{\mu\nu}-ig_2(W^-_{\mu}W^3_{\nu}
      -W^3_{\mu}W^-_{\nu})\right ]\nonumber\\
& & -i\frac{1}{2}(H+v)(\bar t_R\sigma^{\mu\nu} b_L)
      \left
[W^+_{\mu\nu}-ig_2(W^3_{\mu}W^+_{\nu}-W^+_{\mu}W^3_{\nu})\right ],\\
\overline {O}_{tB\Phi}&=&i\frac{1}{\sqrt 2}(H+v)
            (\bar t\sigma^{\mu\nu}\gamma_5 t)B_{\mu\nu},\\
\overline {O}_{\Phi q}^{(1)}&=&(H+v)\partial_{\mu}H \left [\bar
t_L\gamma^{\mu}t_L
          +\bar b_L\gamma^{\mu}b_L\right ] ,\\
\overline {O}_{\Phi q}^{(3)}&=&-\overline {O}_{\Phi
q}^{(1)}+2(H+v)\partial_{\mu}H
        \bar b_L\gamma^{\mu}b_L\nonumber\\
& & -\frac{i}{\sqrt 2}g_2(H+v)^2(W^+_{\mu}\bar t_L\gamma^{\mu}b_L 
                -W^-_{\mu}\bar b_L\gamma^{\mu}t_L) ,
\end{eqnarray}
where $g_1$ and $g_2$ are the gauge couplings for $U(1)$ and
$SU(2)$, $g_Z=2m_Z/v=\sqrt {g_1^2+g_2^2}$.

\appendix{~  Vertices}

The momentum for a fermion is in the direction of a fermion line.
The momentum for a boson is incoming into the vertex.
That is, 
for $\bar t(p_1)-t(p_2)-h(p_3)$, $p_2$ and $p_3$ are incoming and $p_1$ is
outgoing;
for $Z^0_\mu(p_4)-h(p_1)-t(p_3)-\bar t(p_2)$ or 
$\gamma_\mu(p_4)-h(p_1)-t(p_3)-\bar t(p_2)$,
$p_4$, $p_1$, $p_3$ are incoming and $p_2$ is outgoing.
$P_{L,R}$ below are defined as $(1\mp \gamma_5)/2$.

\begin{center}
\large
CP-even Vertices
\end{center}

\begin{eqnarray*}
&&\overline{t}({p_1})-t({p_2})-h({p_3}):\qquad
i\frac{v^2}{\sqrt{2}} \frac{C_{t1}}{\Lambda^2}\\  
&&{Z^0}_\mu ({p_4})-h({p_1})-t({p_3})-\overline{t}({p_2}):\qquad 
i m_Z\frac{C_{t2}}{\Lambda^2} (\gamma_5\gamma_\mu -\gamma_\mu )\\ 
&&\overline{t}({p_1})-t({p_2})-h({p_3}):\qquad
i\frac{1}{2 \sqrt{2}}\frac{C_{Dt}}{\Lambda^2}
(-{p_1}\cdot{p_3}+{p_2}
\cdot{p_3}+ \gamma_5 {p_1}\cdot{p_3}+ \gamma_5 {p_2}\cdot{p_3})\\ 
&&{Z^0}_\mu ({p_4})-h({p_1})-t({p_3})-\overline{t}({p_2}):\qquad 
i\frac{g_2}{12\sqrt{2}c_W}\frac{C_{Dt}}{\Lambda^2}\times \\ 
&&\qquad\qquad (- 3{p_2}_\mu - 3 {p_3}_\mu - 8 \gamma_5 {p_1}_\mu s_W^2 
 +3\gamma_5 {p_2}_\mu - 3 \gamma_5 {p_3}_\mu )\\ 
&&{\gamma}_\mu ({p_4})-h({p_1})-t({p_3})-\overline{t}({p_2}):\qquad 
i\frac{\sqrt{2}}{3}g_2 s_W\frac{C_{Dt}}{\Lambda^2}  
\gamma_5 {p_1}_\mu \\ 
&&{Z^0}_\mu ({p_4})-h({p_1})-t({p_3})-\overline{t}({p_2}):\qquad 
\frac{c_W}{2\sqrt{2}}\frac{C_{tW\Phi}}{\Lambda^2}
\sigma_{\mu\nu}p_4^{\nu}\\ 
&&{\gamma}_\mu ({p_4})-h({p_1})-t({p_3})-\overline{t}({p_2}):\qquad 
\frac{s_W}{2\sqrt{2}}\frac{C_{tW\Phi}}{\Lambda^2}
\sigma_{\mu\nu}p_4^{\nu}\\ 
&&{Z^0}_\mu ({p_4})-h({p_1})-t({p_3})-\overline{t}({p_2}):\qquad 
-\frac{s_W}{2\sqrt{2}}\frac{C_{tB\Phi}}{\Lambda^2}  
\sigma_{\mu\nu}p_4^{\nu}\\  
&&{\gamma}_\mu ({p_4})-h({p_1})-t({p_3})-\overline{t}({p_2}):\qquad  
\frac{c_W}{2\sqrt{2}}\frac{C_{tB\Phi}}{\Lambda^2}
\sigma_{\mu\nu}p_4^{\nu}\\ 
&&{Z^0}_\mu ({p_4})-h({p_1})-t({p_3})-\overline{t}({p_2}):\qquad 
-i 2m_Z\frac{C_{\Phi q}^{(1)}}{\Lambda^2} \gamma_\mu P_L\\ 
&&{Z^0}_\mu ({p_4})-h({p_1})-t({p_3})-\overline{t}({p_2}):\qquad
i 2m_Z\frac{C_{\Phi q}^{(3)}}{\Lambda^2}\gamma_\mu P_L
\end{eqnarray*}
  

\vspace{1cm}
\begin{center}
\large
CP-odd Vertices
\end{center}

\begin{eqnarray*}
&&\overline{t}({p_1})-t({p_2})-h({p_3}):\qquad
i\frac{v^2}{\sqrt{2}} \frac{\overline {C}_{t1}}{\Lambda^2}(i\gamma^5)\\ 
&&\overline{t}({p_1})-t({p_2})-h({p_3}):\qquad
-v\frac{\overline {C}_{t2}}{\Lambda^2}\gamma_\mu P_R p_3^\mu\\ 
&&\overline{t}({p_1})-t({p_2})-h({p_3}):\qquad
\frac{1}{2 \sqrt{2}}\frac{\overline{C}_{Dt}}{\Lambda^2}
(-{p_1}\cdot{p_3}-{p_2}\cdot{p_3}+ 
\gamma_5 {p_1}\cdot{p_3}- \gamma_5{p_2}\cdot{p_3})\\
&&{Z^0}_\mu ({p_4})-h({p_1})-t({p_3})-\overline{t}({p_2}):\qquad
\frac{g_2}{12\sqrt{2}c_W}\frac{\overline{C}_{Dt}}{\Lambda^2}\times\\ 
&& \qquad\quad ( 8 s_W^2 {p_1}_\mu - 3 {p_2}_\mu + 
3 {p_3}_\mu + 3 \gamma_5 {p_2}_\mu + 3 \gamma_5 {p_3}_\mu)\\
&&{\gamma}_\mu ({p_4})-h({p_1})-t({p_3})-\overline{t}({p_2}):\qquad
-\frac{\sqrt{2}}{3}g_2\sin {\theta_w}\frac{\overline{C}_{Dt}}{\Lambda^2}
{p_1}_\mu\\ 
&&{Z^0}_\mu ({p_4})-h({p_1})-t({p_3})-\overline{t}({p_2}):\qquad
\frac{\cos {\theta_w}}{2\sqrt{2}}\frac{\overline{C}_{tW\Phi}}{\Lambda^2}
(i\gamma^5)\sigma_{\mu\nu}p_4^{\nu}\\ 
&&{\gamma}_\mu ({p_4})-h({p_1})-t({p_3})-\overline{t}({p_2}):\qquad
\frac{\sin {\theta_w}}{2\sqrt{2}}\frac{\overline{C}_{tW\Phi}}{\Lambda^2}
(i\gamma^5)\sigma_{\mu\nu}p_4^{\nu}\\ 
&&{Z^0}_\mu ({p_4})-h({p_1})-t({p_3})-\overline{t}({p_2}):\qquad
-\frac{\sin {\theta_w}}{2\sqrt{2}}\frac{\overline{C}_{tB\Phi}}{\Lambda^2}
(i\gamma^5)\sigma_{\mu\nu}p_4^{\nu}\\ 
&&{\gamma}_\mu ({p_4})-h({p_1})-t({p_3})-\overline{t}({p_2}):\qquad
\frac{\cos {\theta_w}}{2\sqrt{2}}\frac{\overline{C}_{tB\Phi}}{\Lambda^2}
(i\gamma^5)\sigma_{\mu\nu}p_4^{\nu}\\
&&\overline{t}({p_1})-t({p_2})-h({p_3}):\qquad
-\frac{\overline{C}_{\Phi q}^{(1)}}{\Lambda^2}v\gamma_{\mu}P_L p_3^{\mu}\\
&&\overline{t}({p_1})-t({p_2})-h({p_3}):\qquad
\frac{\overline{C}_{\Phi q}^{(3)}}{\Lambda^2}v\gamma_{\mu}P_L p_3^{\mu}
\end{eqnarray*}


\end{document}